\documentclass[aps,prc,twocolumn,floatfix]{revtex4}
\usepackage{graphics}
\usepackage{epsfig}

\begin{document}

\title{Track fitting by Kalman Filter method for a prototype cosmic ray muon detector}

\author{Tapasi Ghosh}
\author{Subhasis Chattopadhyay}

\address{Variable Energy Cyclotron Centre, 1/AF BidhanNagar,
                Kolkata-700 064, India.\\
                Tel- 91-33-23371230}

\begin{abstract}
We have developed a track fitting procedure based on Kalman Filter technique for an Iron Calorimeter (ICAL) prototype detector when the detector is flushed with single muon tracks. The relevant track parameters i.e., momentum, direction and charge are reconstructed and analyzed. This paper discusses the design of the prototype detector, its geometry simulation by Geant4, and the detector response with the cosmic ray muons. Finally we show the resolution of reconstructed momenta and also the charge identification efficiency of $\mu^+$ and $\mu^-$ events in the magnetized ICAL.
\end{abstract}
\maketitle


\section{Introduction}
Primary focus of the Iron Calorimeter at India-based Neutrino Observatory (INO)~\cite{ino_report} is to study the interactions involving the atmospheric muon neutrinos and anti-neutrinos. The detector will measure the momentum, direction i.e., tracking, and the sign of the electric charge of muons that are produced by the charged current interactions of the muon neutrinos with the detector material. As a first step, a prototype of ICAL has been already set up at Variable Energy Cyclotron Centre (VECC), Kolkata. The prototype will mainly track cosmic ray muons. Experience with the prototype will be very useful to install much bigger ICAL detector. 
\par 
We are involved to develop a track reconstruction technique for this prototype detector response. Track reconstruction is a process by which one can get an idea of the trajectory of the particle inside the detector. This track reconstruction process can be split into two methods, track finding and track fitting. Different approaches to both track finding and the reconstruction of the initial track parameters are investigated. In present study for the track fitting, ``Kalman Filter''~\cite{cbm_report} technique is utilized. This method is a very useful one which can be used in both track finding and track fitting simultaneously.
\begin{figure}
\centerline{\epsfig{file=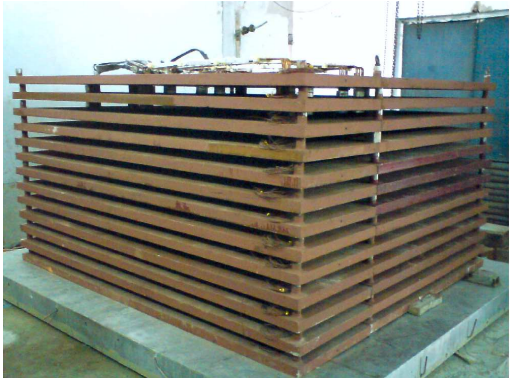,width=5.cm,angle=-90 }}
\caption{INO Iron Calorimeter prototype detector at VECC.}
\label{fig_1}
\end{figure}
\section{Prototype Detector and its Simulation}

\subsection{Detector Geometry}
The ICAL prototype detector~\cite{tapasi_dae08} consisting of 13-layers of iron plates, each having a dimension of $2.5 m \times 2.2 m$, and will provide an effective magnetic field of 1.0 $m^3$ volume. Each iron layer is 5.0 cm thick and inter-spaced with 5.0 cm gap as shown in Fig.~\ref{fig_1}. It is planned that a set of 12 Resistive Plate Chambers (RPCs)~\cite{rpc}, having dimensions of $1.0 m \times 1.0 m$ will be placed inside the iron layers and these RPCs will be of both glass-type and bakelite-type ~\cite{rpc_conf}. The coils which are passing perpendicularly through the iron layers will carry approximate current of 500 ampere. To reduce the heating effect due to this huge amount of current, the coils will be water cooled. The calorimeter will be magnetized with a uniform magnetic field of magnitude 1.0 Tesla (approx.) so that it can distinguish the $\mu^+$ and $\mu^-$ events from their opposite bending natures.

\subsection{Detector Simulation by Geant4}
CERN library based ``detector description and simulation tool'' called GEANT~\cite{geant}, is used to develop the detector geometry as well as to simulate the prototype detector response when the muons pass through it. 

\begin{figure}
\centerline{\epsfig{file=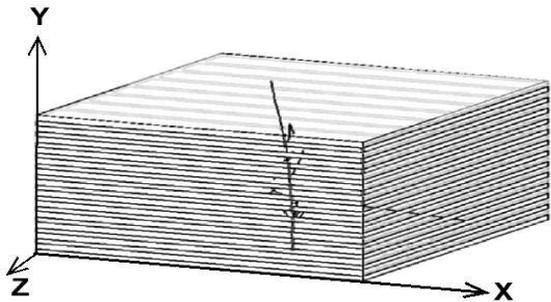,width=7.5cm}}
\caption{\label{fig_2} Simulated prototype detector response while $\mu$ passing through it.} 
\end{figure}

The simulated detector volume consists of iron layers, which are stacked along the y-direction and the pick-up strips are placed along the x and z-directions inside the gap. An uniform magnetic field of magnitude 1 Tesla is applied along the x-direction to magnetize the simulated iron layers. When primary particles i.e., the muons are projected along the y-axis, a clear signature of the bending is visible in Fig.~\ref{fig_2}.
\subsection{Results and Discussions}
The prototype detector will not be able to detect atmospheric neutrinos ($\nu$) due to its limited size and very high cosmic ray muon background on the earth surface.

\begin{figure}
\centerline{\epsfig{file=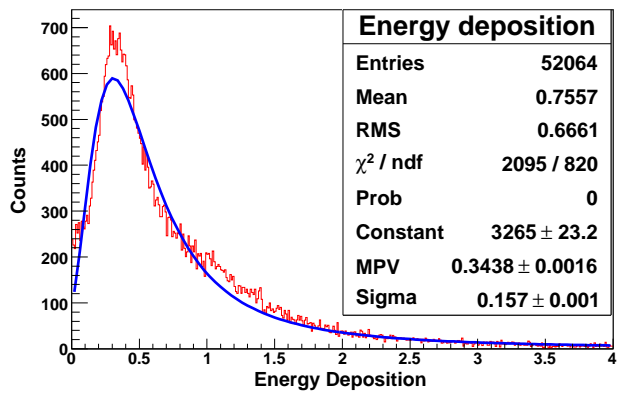,width=9cm}}
\caption{\label{edep} Mean energy deposition of muon inside 2 mm thick RPC gas. The energy deposition spectrum, simulated by GEANT4, has been fitted with Landau spectra.}
\end{figure}

So, high energy atmospheric muon flux  (in the range of 0.5 GeV to 2.0 GeV) on the  earth surface~\cite{cosmic} is used to simulate the detector response.
 The mean value of the energy deposited by $\mu$ in 2.0 mm thick RPC gas is evaluated and the value is approximately 1 keV as shown in Fig.~\ref{edep}.
\section{Track Reconstruction for the prototype detector}
Pattern Recognition is a technique that studies the operation and design  of systems that recognize patterns in data.  In data analysis of HEP experiments, this technique depends on the type of detector used. In tracking detectors, the signals generated by charged particle have to be grouped into track candidates. This is usually called track finding or track search.  In calorimeters, the task of pattern recognition is to group the signals to showers and to compute certain properties of the shower. The importance and feasibility of track and vertex fitting~\cite{kisel} in the more complex environment of experiments with electronic detectors have been recognized over the past three decades ~\cite{fru_book}. Now, we would like to discuss about the Kalman Filter algorithm and its application.

\subsection{Kalman Filter Algorithm}
In the Kalman Filter framework, a track is designated as a set of parameters, called the Kalman state vector $(r)$, which is allowed to change along the particle's path. The procedure starts with a certain initial approximation $r = r_{0}$ and refines the vector $r$, consecutively adding more measurements. The optimum estimation of the state vector is attained after the addition of the last measurement~\cite{dae_07}. The estimation of r is governed by the linear difference equation
\begin{equation}
 r_{k+1} = A_{k}r_{k} + \nu_{k},
\end{equation}
where $A_{k}$ is a linear operator, $\nu_{k}$ is the process noise between $(k-1)_{th}$  and $k_{th}$  measurements. 

If the measurement $ m_{k}$ linearly depends on $r_{k}$, then
\begin{equation}
 m_{k} = H_{k}r_{k} + \eta_{k},
\end{equation}
where $H_{k}$ is the propagator matrix between the measurement space and Kalman state, and $\eta_{k}$ is the measurement noise.

It is assumed that measurement errors $\eta_{i}$ and the process noise $\nu_{j}$ are uncorrelated, unbiased $( \langle \eta_{i}\rangle = \langle\nu_{j}\rangle = 0 )$, and  the covariance matrices $V_{k}$, $Q_{k}$ from these errors are expressed as 
\begin{eqnarray}
< \eta_{i} \cdot \eta^T_{i} > \equiv V_{i},\nonumber\\
< \nu_{j} \cdot \nu^T_{j} > \equiv Q_{j}.
\end{eqnarray}
The conventional Kalman Filter (KF) algorithm consists of following four stages ::\\

$1.$ Initialization Step :: An approximate value of the vector $r_{0}$ is chosen to start with. Its covariance matrix is set to $C_{0}=Inf^2$, where $Inf$ denotes a large positive number.\\

$2.$ Prediction step ::
\begin{eqnarray}
\tilde{r}_{k} & = & A_{k}r_{k-1},\nonumber\\
\tilde{C}_{k} & = & A_{k}C_{k-1}A^T_{k}.
\end{eqnarray}

$3.$ Process noise :: The process noise describes probabilistic deviations of the state vector $r$
\begin{eqnarray}
\hat{r}_{k} & = & \tilde{r}_{k},\nonumber\\
\hat{C}_{k} & = & \tilde{C}_{k} + Q_{k}.
\end{eqnarray}
$4.$ Filtration  step :: In this step $r_{k}$ is updated with the new measurement $m_{k}$ to get the optimal estimate of $r_{k}$ and its covariance matrix
 $C_{k}$
\begin{eqnarray}
K_{k} & = & \hat{C}_{k}H^T_{k} \left(V_{k} + H_{k}\hat {C}_{k}H^T_{k}\right)^{-1},\nonumber\\
r_{k} & = & \hat{r}_{k} + K_{k} \left (m_{k} - H_{k}\hat{r}_{k}\right ),\nonumber\\
C_{k} & = &\hat{C}_{k} - K_{k}H_{k} \hat{C}_{k},\nonumber \\
\chi^2 & =  &\chi^2_{k-1} + \left(m_{k} - H_{k} \hat{r}_{k}\right)^T \left(V_{k} + H_{k} \hat{C}_{k}H^T_{k}\right)^{-1} \nonumber \\ & & \left(m_{k} - H_{k} \hat{r}_{k}\right).
\end{eqnarray}
The matrix $K_{k}$ is known as Kalman gain matrix and the value $\chi^2_{k}$ is the total $\chi^2-$deviation from the measurements $m_{1},....m_{k}$. In brief, in the track fitting algorithm, the track parameters are modified and the reconstructed parameters are obtained from the optimal state vector.  
\subsection{Implementation of Kalman Filter Algorithm and results}
As mentioned earlier, the Kalman Filter algorithm consists of a succession of alternating prediction and filter steps. In this implementation, state vector {\bf r = r(x, y, dx/dz, dy/dz, q/p)} where x and y are for positions, dx/dz and dy/dz are their slopes, and q/p is the charge to momentum ratio. In the prediction step, the current state vector is extrapolated to the next detector plane taking into account multiple scattering and energy loss of the corresponding particle. In the filtration step, the extrapolated state vector is updated by taking a weighted mean with the measurement. This means that after each prediction step it has to be decided which measurement should be used in the subsequent filter step. Conventionally, the measurement which is closest to the prediction is selected for inclusion in the filter.
\begin{figure}
\centerline{\epsfig{file=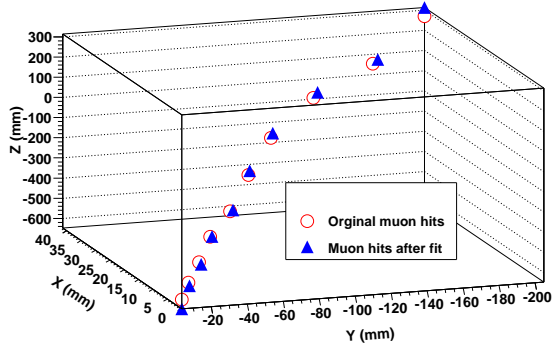,width=8cm}}
\caption{Comparison of hits from a single muon track before and after fit.}\label{track}
\end{figure}
The track fitting starts from a seed track, for our study we consider only single muon tracks and all the tracks are fully-contained inside the simulated prototype detector volume. The initial state vector components are chosen zero, including the initial momentum which reduces biasness. Further more, all the diagonal elements of the state vector covariance matrix are assigned to a large value, whereas the non-diagonal elements are set to zero. The covariance matrix takes care of the multiple scatterings and energy losses due to the passage of the charged particle inside the calorimeter material. At first, the fit proceeds towards downstream direction (from first layer to last layer) and the parameters are transported to the first downstream active plane. This transportation is governed by the $4^{th}$ order Runge-Kutta method which takes care of the bending in magnetic field. A fully-contained single particle track is shown in Fig.~\ref{track} before and after fitting. The track hits will be included in the fit if they corresponds to the least $\chi^2$ value.
 This $\chi^2$ value is the residual between the projected hits and the measured hits. During the propagation of the charge particle inside the matter, we have assumed that the particles do not decay nor have large scattering angle. We also calculate process noise covariance matrix during and after each propagation. The covariance matrix due to measurement noises are set to zero as all the hit points are obtained exactly from Geant4 simulation. Muon deposited approximately 63 MeV energy on an average while traversing one $5.0 cm$ detector plane perpendicularly, as calculated from Bethe-Bloch equation~\cite{pdg}. This energy loss is incorporated to modify the fifth component of state vector i.e., the $q/p$ value as mean energy loss per radiation length. 
\begin{figure}[ht]
\centerline{\epsfig{file=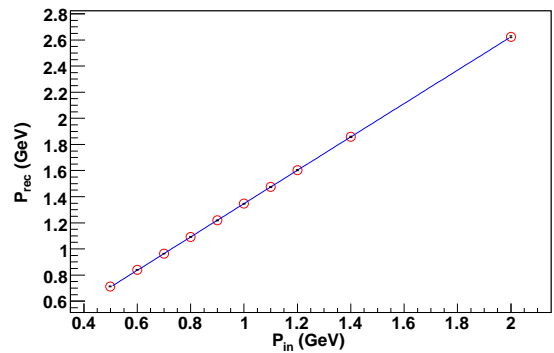,width=8cm}}
\caption{Reconstructed momenta for different incident muon tracks momenta.}
\label{p_prec}
\end{figure}
\begin{figure}[ht]
\centerline{\epsfig{file=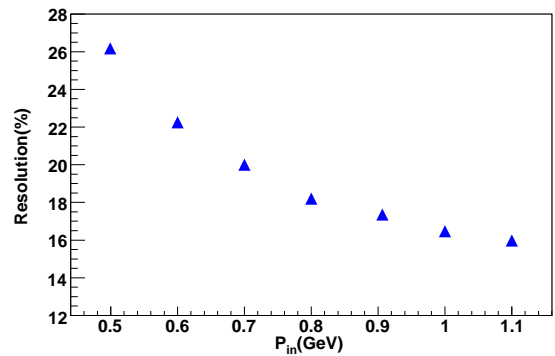,width=8cm}}
\caption{Resolution of reconstructed momenta.}
\label{resolution}
\end{figure}
When the downstream fit ends, its last updated state vector is again used as the starting value for the fit going from downstream to upstream. This upstream fit is performed in the same fashion, starting from same initial covariance matrix as used during downstream fit. Since, Kalman Filter is a recursive algorithm, so we iterate the fit process for 5 iterations for better tuning of the fitted parameters.
\begin{figure}
\centerline{\epsfig{file=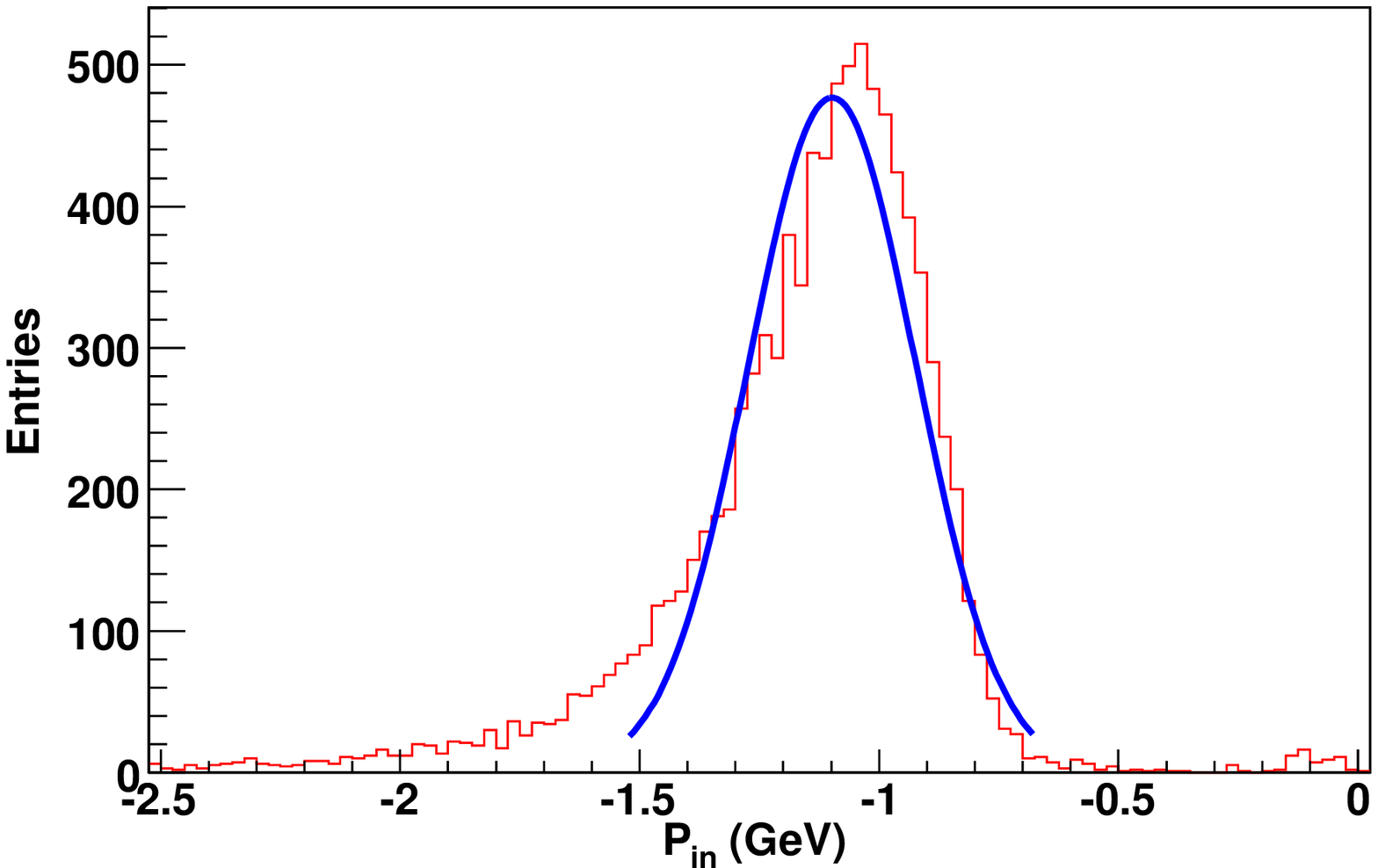,width=8cm}}
\centerline{\epsfig{file=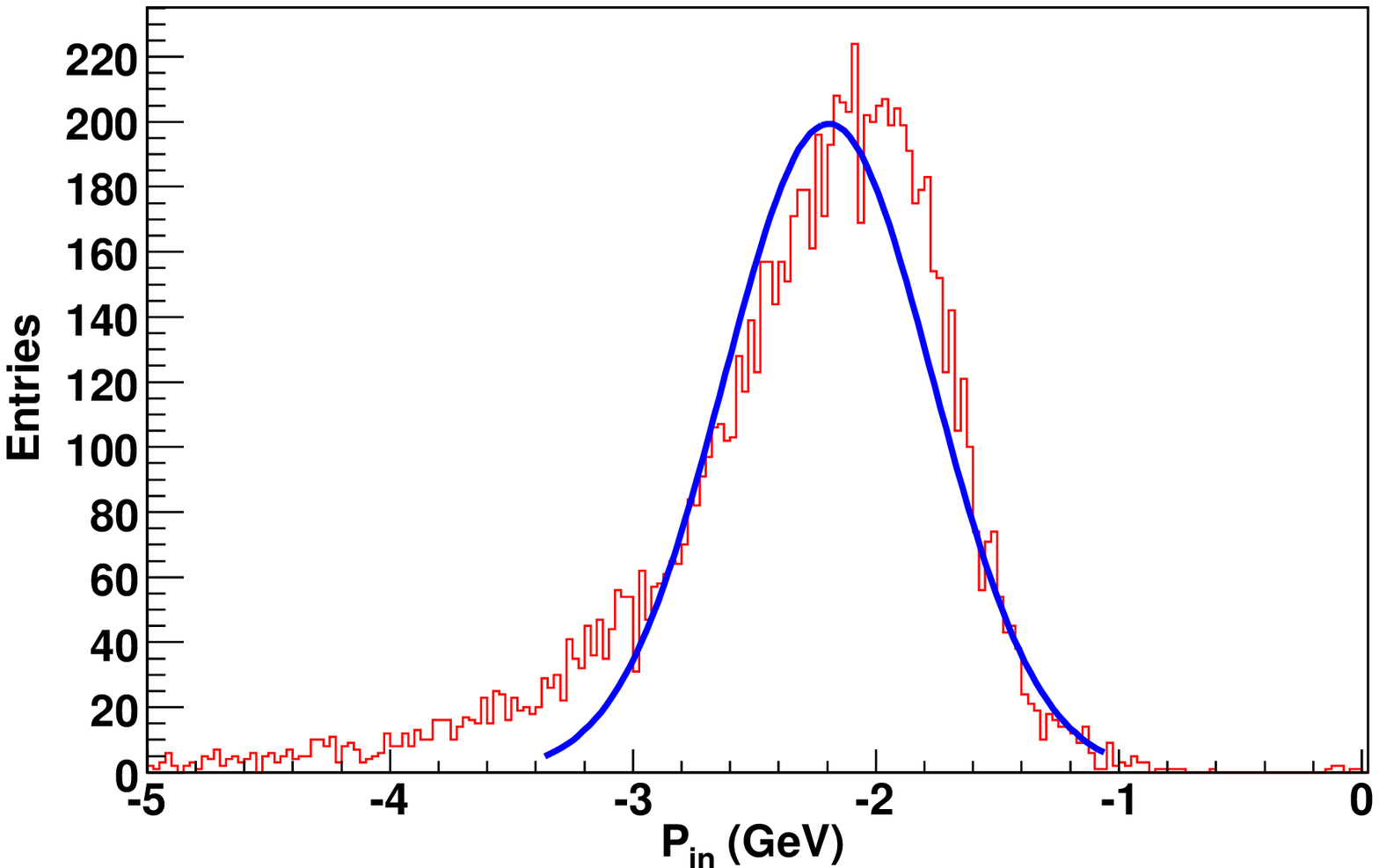,width=8cm}}
\caption{Reconstructed momentum distribution for 1 GeV and 2 GeV muon momenta.}
\label{reco_mom}
\end{figure}

The linearity of the reconstructed momenta with the incident momenta of the reconstructed tracks are shown in Fig.~\ref{p_prec}. Even though the linearity is found to be very good, the slope seems to have deviated slightly from 1 presumably due to partially contained tracks and fluctuations in energy deposition in the calorimeter material. For subsequent analysis, we have worked with calibrated reconstructed momenta to make the slope close to unity. 
Performance of the fitting is characterized by the value of reconstructed momentum resolution as shown in Fig.~\ref{resolution}.
Now, 
\begin{eqnarray}
{\rm resolution} \ ( \%)  \ =  \ {\frac { \sigma }{ M}} \ \times \ 100.
\end{eqnarray}
Where $\sigma$ is the HWFM of the gaussian and M is the mean value obtained from the fit as shown in Fig.~\ref{reco_mom}. For fully-contained muon track i.e., the track whose end hit point is inside the detector volume, we can estimate its initial energy very well. For this prototype, the muon tracks are fully contained upto 1.1 GeV, although the program is sensitive enough to reconstruct the partially contained events as shown in Fig.~\ref{p_prec}. However, for momenta beyond 1.2 GeV, the resolutions have been found to get worsened. The main reason is that for higher energy, the particles go beyond the 13 detector layers.

\begin{figure}[ht]
\centerline{\epsfig{file=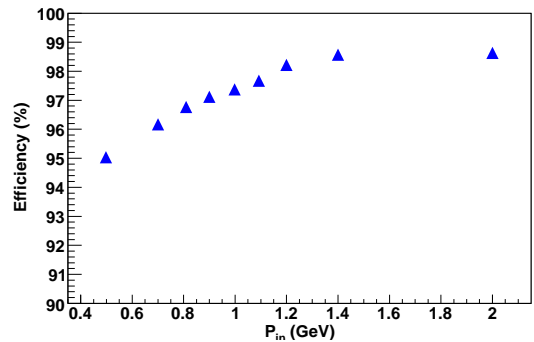,width=7.7cm}}
\caption{Charge identification efficiency for various incident momenta of muon events.}
\label{charge}
\end{figure}

The charge identification efficiency is calculated for $\mu^+$ and $\mu^-$ events from their opposite bendings in magnetic field and it is observed that the efficiency is upto 99 $\%$ for 2 GeV muon as shown in Fig.~\ref{charge}. The significance of this charge separation is that in future this program will be able to distinguish neutrino and anti-neutrino events which produce $\mu^-$ and $\mu^+$ respectively by interacting with the atoms of detector material.
\section{Summary}
We have presented a track fitting simulation procedure for a cosmic ray prototype detector including all the physics associated with the passage of a charged particle through matter. We have only assumed the single particle muon track per event incident on the detector. It is found that the program can estimate both fully contained and partially contained events with good resolution. We believe that by increasing the no of layers, the resolution of fitting can be improved. Finally, the procedure can prominently distinguish opposite charges of the $\mu^+$ and $\mu^-$ events.
\section{Acknowledgments}
We would like to thank Dr. Ivan Kisel, Prof. Sudeb Bhattacharya, Prof. Naba Mandal, Dr. D. Indumathi and Dr. Gobinda Majumdar for their all kinds of help, support and suggestions.

\end{document}